\documentclass{aa}
\usepackage{graphicx}
\usepackage{color}
\usepackage{graphicx}
\usepackage{lscape}
\usepackage[varg]{txfonts}
\usepackage{amssymb}
\usepackage{stmaryrd}
\usepackage{url}
\usepackage{xspace}
\usepackage{siunitx}
\usepackage{textcomp}

\usepackage{natbib,twoopt}
\usepackage[breaklinks=true]{hyperref} 
\bibpunct{(}{)}{;}{a}{}{,} 
\makeatletter
\newcommandtwoopt{\citeads}[3][][]{\href{http://adsabs.harvard.edu/abs/#3}%
{\def\hyper@linkstart##1##2{}%
\let\hyper@linkend\@empty\citealp[#1][#2]{#3}}}
\newcommandtwoopt{\citepads}[3][][]{\href{http://adsabs.harvard.edu/abs/#3}%
{\def\hyper@linkstart##1##2{}%
\let\hyper@linkend\@empty\citep[#1][#2]{#3}}}
\newcommandtwoopt{\citetads}[3][][]{\href{http://adsabs.harvard.edu/abs/#3}%
{\def\hyper@linkstart##1##2{}%
\let\hyper@linkend\@empty\citet[#1][#2]{#3}}}
\newcommandtwoopt{\citeyearads}[3][][]%
{\href{http://adsabs.harvard.edu/abs/#3}
{\def\hyper@linkstart##1##2{}%
\let\hyper@linkend\@empty\citeyear[#1][#2]{#3}}}
\makeatother

\newcommand{\hen}{\object{Hen\,2$-$428}\xspace}
\newcounter{Rco}
\newcommand{\Ionst}[1]{\setcounter{Rco}{#1}\Roman{Rco}}
\newcommand{\Ion}[2]{\mbox{#1\,{\scriptsize\Ionst{#2}}}}
\newcommand{\Ionw}[3]{\mbox{#1\,{\scriptsize\Ionst{#2}}~$\lambda\,#3$\,\AA}}

\newcommand{\Ionww}[3]{\mbox{#1\,{\scriptsize\Ionst{#2}}~$\lambda\lambda\,#3$\,\AA}}

\newcommand{\loggw}[1]{\mbox{$\log g\hspace{-0.5mm} =\hspace{-0.5mm}  #1$}}

\newcommand{\se}[1]{\mbox{Sect.\,\ref{#1}}}

\newcommand{\Teff}{\mbox{$T_\mathrm{eff}$}\xspace}

\newcommand{\ebv}{$E_\mathrm{B-V}$\xspace}

\newcommand{\Lsol}{$L_\odot$}
\newcommand{\Msol}{$M_\odot$}
\newcommand{\Rsol}{$R_\odot$}

\begin{document}

\title{An in-depth reanalysis of the alleged type Ia supernova progenitor Henize 2-428}

\author{N. Reindl\inst{1}
  \and V. Schaffenroth\inst{1}
  \and M. M. Miller Bertolami\inst{2}$^{,}$\inst{3}
  \and S. Geier\inst{1}
  \and N. L. Finch\inst{4}
  \and M. A. Barstow\inst{4}
  \and\\ S. L. Casewell\inst{4}
  \and S. Taubenberger\inst{5}
}

\institute{Institute for Physics and Astronomy, University of Potsdam,
  Karl-Liebknecht-Str. 24/25, D-14476 Potsdam, Germany\\ \email{nreindl885@gmail.com}
  \and Instituto de Astrof\'{i}sica de La Plata, UNLP-CONICET, La Plata, 1900 Buenos Aires, Argentina
  \and Facultad de Ciencias Astron\'{o}micas y Geof\'{i}sicas, UNLP, Buenos Aires, Argentina Paseo del Bosque s/n, FWA, B1900 La Plata, Buenos Aires, Argentina
  \and Department of Physics and Astronomy, University of Leicester, University Road, Leicester LE1 7RH, UK
  \and Max Planck Institut f\"ur Astrophysik, Karl-Schwarzschild-Straße 1, 85748 Garching, Germany}

\date{Received 7 April 2020 / Accepted 14 May 2020}

\abstract{The nucleus of the planetary nebula Hen\,2-428 is a short
  orbital-period (4.2\,h), double-lined spectroscopic binary, whose status as a
  potential supernova type Ia progenitor has raised some controversy in the literature.}
  {With the aim of resolving this debate, we carried out an in-depth reanalysis of the system.}
  {Our approach combines a refined wavelength calibration, thorough
    line-identifications, improved radial-velocity measurements, non-LTE spectral modeling, as well as
    multi-band light-curve fitting. Our results are then discussed in view of state-of-the-art stellar evolutionary models.}
  {Besides systematic zero-point shifts in the wavelength calibration of the
  OSIRIS spectra which were also used in the previous analysis of the system, we found
  that the spectra are contaminated with diffuse interstellar bands. Our Voigt-profile radial
  velocity fitting method, which considers the additional absorption of these diffuse interstellar bands, 
  reveals significantly lower masses ($M_1=0.66\pm0.11$\,\Msol\ and
  $M_2=0.42\pm0.07$\,\Msol) than previously reported and a mass ratio that is 
  clearly below unity. Our spectral and light curve analyses lead
  to consistent results, however, we find higher effective temperatures and
  smaller radii than previously reported. Moreover, we find that the red-excess
  that was reported before to prove to be a mere artifact of an outdated reddening law
  that was applied.}
{Our work shows that blends of He\,{\sc ii}\,$\lambda\,5412$\,\AA\ with
  diffuse interstellar bands have led to an
  overestimation of the previously reported dynamical masses of
    Hen\,2-428. The merging event of Hen\,2-428 will not be recognised as a
    supernova type Ia, but most likely leads to the formation of a H-deficient star. We suggest that the system was formed via a first stable mass
    transfer episode, followed by common envelope evolution, and it is now composed of a
    post-early asymptotic giant branch star and a reheated He-core white dwarf.}

\maketitle

%

\section{Introduction}
\label{sect:intro} 

The detection and analysis of compact binary systems is fundamental to various areas
of astrophysics \citep{Jones2020}. Binary interactions are thought to play a key role in the
shaping of planetary nebulae (PNe, \citealt{DeMarco2009, Jones2019}) and are needed to explain the
formation of diverse objects, such as hot sub\-dwarf stars, extremely low mass
white dwarfs \citep{Paczynski1976, Webbink1984, IbenTutukov1986},
or post-red giant branch (RGB) central stars of planetary nebulae (CSPNe, \citealt{Hall+2013, Hillwig+2017}).
Compact binaries are crucial to understand common envelope (CE) evolution and
they serve as important tests for general relativity as very close binary systems (periods 
of less than a few hours) emit considerable amounts of gravitational radiation
\citep{WeisbergTaylor2005, Burdge+2019}.\\
The emission of gravitational waves in very close white dwarf binary systems
leads to a shrinkage of their orbits resulting in mass transfer between the white dwarfs
or even the merger of the white dwarfs. The ultimate fate
of these systems depends on their total mass as well as the mass ratio,
$q=M_2/M_1$, and whether mass transfer remains dynamically stable or 
not \citep{Shen2015}. The outcomes of such interaction have been proposed to lead to the
formation of exotic objects showing He-dominated atmospheres such as R Coronae Borealis stars (RCB), extreme
helium (EHe) stars, He-rich hot subdwarf O (He-sdO) stars, or O(He) stars
\citep{Webbink1984, Iben1984, SaioJeffery2002, justham2011, zhangetal2012a, zhangetal2012b,
  Zhang2014, Reindletal2014b}. Also stars with C/O-dominated
atmospheres such as the very hot white dwarfs H1504+65 and RXJ0439.8$-$6809
\citep{WernerRauch2015}, WO-type central stars \citep{Gvaramadze+2019}, or hot
DQ white dwarfs \citep{Kawka+2020} have been proposed to
be the outcome of such mergers.\\
For sufficiently high mass progenitors the merger of the two white dwarfs can
also lead to Type Ia supernovae (SN\,Ia) or faint thermonuclear supernovae
(SN\,.Ia), which reach only one-tenth of the brightness of a SN\,Ia.
This may occur via the so-called double-degenerate channel in which the
resulting merger has a mass near the Chandrasekhar
limit \citep{Iben1984, Webbink1984}, but various other evolutionary
pathways for the double degenerate SN\,Ia channel have been proposed for which the progenitor
systems may also have sub-Chandrasekhar masses. These include 
the double-detonation mechanism \citep{WoosleyWeaver1994, Fink+2007, Fink+2010, Liu+2018, Shen+2018}, 
the violent merger model \citep{Pakmor+2011, Pakmor+2013, Liu+2016}, or the core degenerate channel
\citep{SparksStecher1974, KashiSoker2011}.\\
The detection of progenitor systems for the double-degenerate SN\,Ia model is
extremely challenging as recently demonstrated by
\cite{Rebassa-Mansergas+2019}, who predict an observational probability only of
the order $10^{-5}$ for finding double white dwarf SN\,Ia progenitors in
our Galaxy with current telescopes. Large observational efforts to
search for double-degenerate SN\,Ia progenitor systems amongst double white
dwarfs or white dwarf and pre-white dwarf (hot subdwarf) systems 
\citep{Napiwotzkietal2001, Geieretal2011a, Breedt+2017, Napiwotzki+2019} have
revealed some progenitor candidates \citep{Maoz+2014}, but none of them has been confirmed
unambiguously and robustly. The only exception might be the
binary system residing in the planetary nebula \hen, which is subject of this
paper.\\
\hen was discovered by \cite{Henize1976} and a first hint of the binarity of
its nucleus was suggested by \cite{Rodriguezetal2001} based on the discovery of a red-excess
emission. The non-ambiguous evidence that \hen hosts a binary central star,
was only delivered by \cite{SG+2015} (hereafter SG+15). They made the stunning discovery that
the system is a double-lined spectroscopic binary system composed of two hot pre-white
dwarfs. Fitting Gaussian profiles to the double lined and time variable \Ionw{He}{2}{5412}
line they found the radial velocity (RV) semi-amplitudes of both stars to be the
same ($206\pm8$\,km/s and $206\pm12$\,km/s). In addition, they derived a
photometric period of 4.2\,hours and showed that the light curves can be
reproduced assuming an over-contact system seen at an inclination angle of
$i=64.7\pm1.4$\textdegree. From this they derived dynamical masses of $0.88\pm0.13$\Msol\ for both
stars, and concluded that the system is composed of two hot pre-white dwarfs
with a combined mass higher than the Chandrasekar limit which will merge
within 700 million years triggering a SN\,Ia.\\
This scenario has since been challenged by \citet{GB+2016}, who criticized the
strong mismatch between the luminosities and radii of both pre-white dwarf
components as derived by SG+15 with the predictions from single-star stellar evolution
models \citep{Bloecker1991, Bloecker1993, Renedo+2010}. In addition,
\cite{GB+2016} suggested that the variable 
\Ionw{He}{2}{5412} line might instead be a superposition of an absorption line
plus an emission line, possibly arising from the nebula, the irradiated
photosphere of a close companion, or a stellar wind. Since this would question 
the dynamical masses derived by SG+15, \cite{GB+2016} repeated the light curve
fitting and showed that the light curves of \hen may also be fitted well by
assuming an over-contact binary system that consists of two lower mass (i.e.,
masses of 0.47\,\Msol\ and 0.48\,\Msol) stars. Thus, they concluded that the claim 
that \hen provides observational evidence for the double degenerate scenario
for SN\,Ia is premature.\\
Given the potential importance of \hen as a unique laboratory to study the
double degenerate merger scenario, it is highly desirable to resolve this
debate. This is the goal of this work. The paper is organized as
follows. In \se{sect:obs}, we give an overview of the available observations,
and provide a detailed description of continuum and line contributions to
the spectra (\se{sect:spectra}). 
In \se{sect:RVs}, we examine the wavelength calibration accuracy and perform an improved RV analysis.
After that we carry out a non-LTE spectral analysis to derive
atmospheric parameters (\se{sect:atmo}) and perform multi-band light-curve
fits (\se{sect:photo}). The dynamical masses are presented in \se{sect:masses}
along with a discussion of the evolutionary status of the system.
We summarize and present our conclusions in \se{sect:discussion}.

\section{Observations}
\label{sect:obs}

\subsection{Photometry}
\label{sect:photometry}

SG+15 obtained time-resolved i-band (effective wavelength
$0.44\,\mathrm{\mu m}$) photometry with the MERcator Optical
Photometric ImagEr (MEROPE, \citealt{Davignon+2004}) on the
\textit{Mercator} telescope on La Palma on 28 and 30 August 2009,
and on 2 September 2009. Another i-band time-series was obtained
by them on 2 August 2013 with the Wide Field Camera at the 2.5m
\textit{Isaac Newton} Telescope (INT) as well as a Johnson
B-band (effective wavelength $0.78\,\mathrm{\mu m}$) time-series
with the South African Astronomical Observatory (SAAO) 1\,m
telescope on 11 July 2013. In addition, we acquired Asteroid
Terrestrial-impact Last Alert System (ATLAS, \citealt{Tonry+2018})
c- and o-band light curves (effective wavelengths
$0.53\,\mathrm{\mu m}$ and $0.68\,\mathrm{\mu m}$, respectively)
of \hen.

\subsection{Spectroscopy}
\label{sect:spectra}

Low-resolution spectroscopy of \hen was obtained by \cite{Rodriguezetal2001} using
the Intermediate Dispersion Spectrograph (IDS) at the INT. These observations have a spectral resolution of
$\approx 8$\,\AA\ and cover the wavelength range of $3500-9000\,\AA$.\\
SG+15 obtained four observations with the FOcal Reducer/low
dispersion Spectrograph 2 (FORS2) mounted on the Unit Telescope 1 (UT1)
of the ESO Very Large Telescope (VLT) array (ProgIDs:
085.D-0629(A), 089.D-0453(A)). The observations were obtained in 2010 and 2012 using
the 1200G grism (spectral resolution of $\approx 3$\,\AA, resolving power
$R=1605$). We downloaded these observations from the ESO archive and reduced
them using standard IRAF procedures.\\
The most useful set of observations (15 exposures in total,
Table~\ref{tab:obs}) was obtained at the Gran Telescopio
Canarias (GTC) using the Optical System for Imaging and low Resolution
Integrated Spectroscopy (OSIRIS) with the R2000B grating (ProgID: GTC41-13A).
The spectra ($R=2165$) with a mean exposure time of 868\,s cover the full
orbital period of the system and were used by SG+15 to derive the
RV curves of the system. The signal-to-noise ratio (S/N) of these
observations is similar to the ones of the FORS2 observations ($\approx 70$ at
4600\,\AA), but they have a higher resolution (2\,\AA\ instead of 3\,\AA). We 
downloaded the OSIRIS observations that were reduced by SG+15 from
the GTC Public Archive.\\  
Additionally, we obtained observations using the UV-Visual
high-resolution Echelle Spectrograph (UVES) mounted at the 8.2\,m Kueyen (UT2)
telescope (ProgID: 295.D-5032(A)). The poor signal to noise (S/N$\approx$5) of these
spectra, however, does not allow the identification of photospheric lines, thus we discarded these 
observations from our analysis of the central stars.

\begin{figure*}[ht]
\centering
\includegraphics[width=\textwidth]{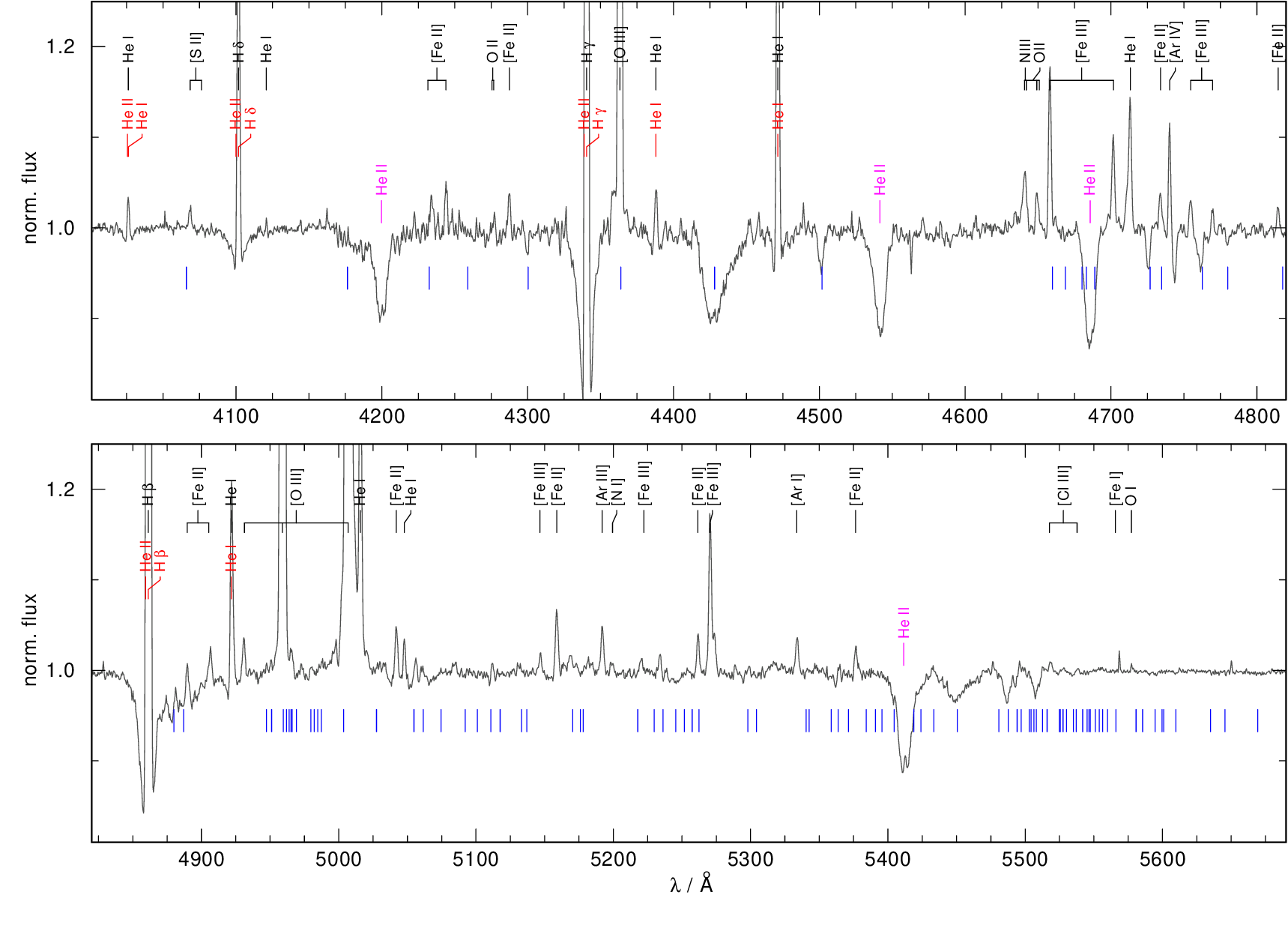}
\caption{Co-added and normalized OSIRIS spectrum. The locations of known
  diffuse interstellar bands (blue), nebular (black), and photospheric lines
  (red) are marked. Photospheric lines used in for
the RV analysis are marked in magenta.}
\label{fig:osiris}
\end{figure*}

The spectra of \hen are a complex superposition of nebular, photospheric,
interstellar, and circumstellar contributions. The continuum flux is
noticeably affected by interstellar and circumstellar
reddening and - as claimed by \cite{Rodriguezetal2001} - possibly by a late
type companion that causes a red-excess. For the further analysis it is
crucial to first disentangle and check these various contributions.

\subsubsection{Nebular contributions}
\label{sect:neblines}

The nebula only contributes a negligible fraction to the continuum flux,
for instance \cite{Rodriguezetal2001} estimated that the strongest source of nebular
continuum (the recombination continuum of \Ion{H}{1}) may only account for a
few percent to the total flux in the optical. Much more prominent are the
nebular emission lines, which are labeled in black in Fig.~\ref{fig:osiris}
where we show the coadded OSIRIS spectrum.\\
For the nebular line identifications we made use of the nebular line list for \hen provided in
\cite{Rodriguezetal2001}, published line lists of other PNe \citep{Zhang+2012, Corradi+2015}, as well as The Atomic Line List
v2.05b21\footnote{\url{https://www.pa.uky.edu/~peter/newpage/}}
\citep{VanHoof2018}. Thanks to the higher resolution of the OSIRIS 
spectra compared to the IDS spectra used in the nebular analysis by
\cite{Rodriguezetal2001}, we found in addition also collisionally excited lines of [\Ion{Cl}{3}], [\Ion{Ar}{1}],
[\Ion{Ar}{4}], [\Ion{Fe}{1}], [\Ion{Fe}{2}], and [\Ion{Fe}{3}] as well as
optical recombination lines of \Ion{N}{3} and \Ion{O}{2}. The latter could be
blended with photospheric lines, but since they do not
vary over the orbital period, we conclude that these lines mainly originate from the
nebula.\\
With regard to the concept of \citet{GB+2016}, that the small reversals in the
cores of the \Ion{He}{2} lines might originate from nebular
emission lines, we note that in this case the nebular line flux of
\Ionw{He}{2}{4686} should be about one order of magnitude higher than that of the remaining 
\Ion{He}{2} lines (e.g., \citealt{Zhang+2012}). In addition, it was already reported by 
\cite{Tylenda+1994} and SG+15 that \hen does not show the \Ionw{He}{2}{4686}
nebular line. Therefore, the presence of the \Ionww{He}{2}{4200, 4542, 5412}
nebular lines, which are much weaker, can be excluded as well. We also note
that no \Ion{He}{2} emission lines can be detected in the UVES observations.

\subsubsection{Photospheric contributions}
\label{sect:photolines} 

The spectra show photospheric absorption lines of \Ionww{H}{1}{4102, 4340, 4861} 
which are blended with photospheric absorption lines of \Ionww{He}{2}{4100, 4339, 4859},
as well as \Ionww{He}{1}{4026, 4388, 4472, 4922} (marked in red in
Fig.~\ref{fig:osiris}, \Ionww{He}{1}{4026} is blended with the weaker \Ionww{He}{2}{4026}).
All of these lines are blended with nebular lines, i.e. they show photospheric
absorption wings, while the line cores exhibit either \Ion{H}{1} or
\Ion{He}{1} nebular emission lines. The only photospheric lines, which are not blended with nebular lines are
\Ionww{He}{2}{4200, 4542, 4686, 5412} (marked in magenta in Fig.~\ref{fig:osiris}).\\
We also would like to comment here on the idea of \cite{GB+2016}
that the \Ionw{He}{2}{5412} line might be a superposition of a single absorption line
plus an emission line. Compared to synthetic spectra for hot (pre-)white
dwarfs, all \Ion{He}{2} lines in the spectra of \hen are at the same time
unusually broad and deep.
This could, in principle, be explained by a pure He atmosphere of a very fast rotating
star. However, then the observed absorption wings of the Balmer lines should
be much weaker. Thus, we conclude that the \Ion{He}{2} lines are indeed double-lined
and stem from the photospheres of the two hot stars as reported by SG+15. This
is also perfectly supported by the RV analysis (\se{sect:RVs}).\\
The fluxes of the two hot stars constitute the dominant contribution to the
continuum flux, whose shape is, however, altered by
reddening which we will discuss in the next section.

\begin{figure}[t]
\centering
\includegraphics[width=\columnwidth]{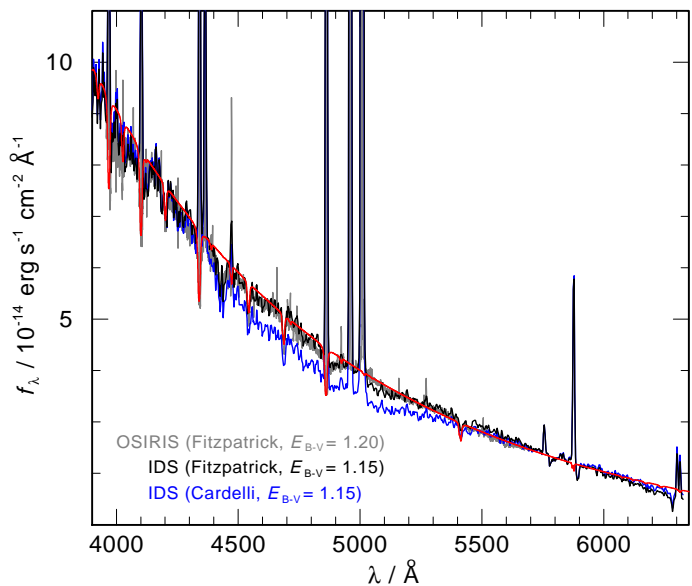}
\caption{Determination of the reddening. The OSIRIS spectrum \#2 de-reddend
  with the Fitzpatrick reddening law (gray) and the IDS spectrum de-reddened
  once with the Fitzpatrick reddening law (black) and once with the Cardelli
  reddening law (blue) are compared to our best fit TMAP model (red).}
\label{fig:ebv}
\end{figure}

\subsubsection{Interstellar and circumstellar contributions}
\label{sect:reddening}

The determination of the reddening of the observations is important to investigate
the nature of the claimed red-excess by \cite{Rodriguezetal2001}, which could
have a noticeable impact on the RV, light curve, and spectral analysis.
This is because a cool companion might leave behind spectral features
contaminating the spectrum and add an additional continuum
light to both the spectroscopic and photometric observations. Furthermore, the
knowledge of the reddening is also essential for the distance determination.\\
We determined the reddening by de-reddening the IDS and
OSIRIS observations for different values of \ebv with the reddening law of
\cite{Fitzpatrick1999} until a good agreement with our best fit model spectrum (see \se{sect:atmo})
was found. In Fig.~\ref{fig:ebv}, our best fit model
is shown in red, the de-reddened OSIRIS spectrum \#2
in gray, and the de-reddened IDS spectrum in black. For the IDS spectrum we
find \mbox{\ebv$=1.15\pm0.05$\,mag} corresponding to \mbox{$A_V=3.57\pm0.16$\,mag}
(assuming \mbox{$R_V=3.1$}), while the OSIRS observations suggest a slightly
higher value of \mbox{\ebv$=1.20\pm0.05$\,mag}. 
Using the nebula line ratio of H$\alpha$/H$\beta$ and the reddening law of
\cite{Cardelli+1989}, \cite{Rodriguezetal2001} found \mbox{$A_V=2.96\pm0.34$\,mag},
which is about 20\% smaller than the values derived by us\footnote{We note that
differences up to 50\% in \ebv as derived from different spectra or nebula
lines have also been noticed in the CSPN SAO\,245567 \citep{Arkhipova2013, Reindletal2014a}.}.
\\
In Fig.~\ref{fig:ebv} we also show in blue the IDS spectrum de-reddened with
the reddening law of \cite{Cardelli+1989} instead of the 
\cite{Fitzpatrick1999} law. This causes a clearly visible depression of the observed flux
from 4400 to 5400\,\AA, which reaches its maximum deviation from the model
spectrum around 4900\,\AA. The Cardelli reddening law is considered as outdated,
i.e. based on fits to the location of the blue tip of the stellar locus in
various SDSS fields, \cite{Schlafly+2010} report that the Fitzpatrick reddening law is 
clearly favored over Cardelli. We therefore conclude that the red excess
claimed by \cite{Rodriguezetal2001} is merely a consequence of the reddening law
used, which makes it appear as if there is an increased continuum emission
red-wards of about 5000\,\AA. This also implies that \hen has likely no 
late-type companion, at least none that is noticeable in the optical wavelength
range.\\
\hen is located at a low galactic latitude ($b=2.48$\textdegree) and embedded
in the galactic disk, therefore the relatively high extinction towards this
source is not surprising. The 3D Reddening Map of interstellar dust by
\cite{Lallement+2018}\footnote{https://stilism.obspm.fr/} extends to 2.63\,kpc
in the direction of \hen (which is close to the distance given by
\cite{Frew+2016}, who derived $2.72\pm0.86$\,kpc using the H$\alpha$ surface
brightness–radius relation) and predictes \mbox{\ebv$=0.81\pm0.05$}. Thus,
about one third of the reddening towards \hen might be circumstellar and
caused by the compact nebula.\\

\begin{table*}[ht]
\caption{OSIRIS observations of \hen. The Heliocentric Julian Day at middle of
  the exposure, exposure times, and RVs as measured from the different
  Balmer lines (indicating the drift of the zero-point in course of
    the observing run) are listed.}
\label{tab:obs}
\begin{tabular}{l l l c c c c c} 
Nr.  & ID  &  HJD$_\mathrm{middle}$ & $t_\mathrm{exp}$\,[s] & H$\delta$\,[km/s] & H$\gamma$\,[km/s] & H$\beta$\,[km/s] & H$\delta, \gamma, \beta$\,[km/s] \\
\noalign{\smallskip}
\hline 
\noalign{\smallskip}
\#1	 & 0000411146  & 2456516.44874	 & 868 & $65\pm3$ & $65\pm1$ & $69\pm1 $ & $66\pm1$ \\
\#2	 & 0000411147  & 2456516.45964	 & 868 & $67\pm5$ & $67\pm2$ & $69\pm2 $ & $68\pm6$ \\
\#3	 & 0000411148  & 2456516.47053	 & 868 & $62\pm6$ & $62\pm2$ & $65\pm1 $ & $63\pm2$ \\
\#4	 & 0000411149  & 2456516.48142	 & 868 & $56\pm2$ & $55\pm2$ & $59\pm2 $ & $56\pm4$ \\
\#5	 & 0000411150  & 2456516.49231	 & 868 & $53\pm3$ & $54\pm1$ & $58\pm2 $ & $55\pm5$ \\
\#6	 & 0000411152  & 2456516.52202	 & 650 & $34\pm2$ & $35\pm1$ & $42\pm2 $ & $37\pm5$ \\
\#7	 & 0000411161  & 2456516.55371	 & 868 & $30\pm3$ & $33\pm1$ & $39\pm4 $ & $34\pm4$ \\
\#8	 & 0000411162  & 2456516.56460	 & 868 & $28\pm2$ & $28\pm2$ & $35\pm2 $ & $30\pm2$ \\
\#9	 & 0000411163  & 2456516.57549	 & 868 & $29\pm3$ & $28\pm3$ & $36\pm2 $ & $31\pm5$ \\
\#10	 & 0000411164  & 2456516.58638	 & 868 & $17\pm2$ & $17\pm3$ & $23\pm2 $ & $19\pm6$ \\
\#11	 & 0000411165  & 2456516.59728	 & 868 & $14\pm3$ & $14\pm3$ & $20\pm2$  & $16\pm5$ \\
\#12	 & 0000411166  & 2456516.60817	 & 868 & $12\pm3$ & $11\pm4$ & $20\pm4$  & $14\pm6$ \\
\#13	 & 0000411167  & 2456516.61906	 & 868 & $20\pm2$ & $17\pm3$ & $21\pm2$  & $19\pm4$ \\
\#14	 & 0000411168  & 2456516.62995	 & 868 & $20\pm3$ & $16\pm2$ & $20\pm1$  & $19\pm5$ \\
\#15	 & 0000411169  & 2456516.64084	 & 868 & $24\pm3$ & $20\pm2$ & $22\pm3$  & $22\pm5$ \\
\noalign{\smallskip}
\hline
\end{tabular}
\end{table*}

Interstellar and circumstellar contributions, however, do not only leave a noticeable impact the continuum flux.
The spectra of \hen also exhibit numerous
additional absorption lines, which we all identify as absorptions caused by diffuse interstellar bands (DIBs). 
These absorption features, often seen in highly reddened stars, originate
in the interstellar medium (ISM) and are typically broader than expected
from the Doppler broadening of turbulent gas motions in the ISM
\citep{Jenniskens+Desert1994}. DIBs are widely assumed to be caused by large
molecules (e.g., C$_{60}^+$ \citealt{Campbell+2015}), however, not all DIBs have
yet been conclusively identified. In the
wavelength range from $4\,000$ to $10\,000$\,\AA, there are several classes of
molecules considered to be possible DIB absorbers and which may produce a few
strong bands along with a much larger array of weaker bands
(\citealt{Hobbs+2008} and references therein).\\
The blue bars in Fig.~\ref{fig:osiris} mark the locations of DIBs identified in the high-resolution,
high S/N spectrum of HD\,204827 by \cite{Hobbs+2008}. We note that
due to the lower resolution and lower S/N of the OSIRIS observations, only
relatively strong DIBs are visible. The strength of the most prominent DIB at 4430\,\AA\
resembles the strengths of the photospheric \Ion{He}{2} lines, but other
strong DIBs at 4501.79, 4726.83, 4762.61, 5450.62, 5487.69, 5525.48\,\AA\AA\ are
clearly visible as well. HD\,204827 has a very similar reddening (\mbox{\ebv$=1.11$},
\citealt{Hobbs+2008}) compared to \hen (\mbox{\ebv$=1.15$}, see above). 
Since the equivalent width of DIBs is correlated to
the value of \ebv (e.g., \citealt{KosZwitter2013, Krelowski+2019}),
one can expect that the DIBs in \hen are of similar
strengths to what is observed in \object{HD\,204827}.
A quite crucial point that now becomes obvious when looking at Fig.~\ref{fig:osiris} is
that three of the four \Ion{He}{2} lines are blended with
DIBs. \Ionw{He}{2}{4686} and \Ionw{He}{2}{5412} are blended with
three and four weaker DIBs, respectively. Bluewards (at about 4176\,\AA) of \Ionw{He}{2}{4200} a
relatively broad and strong DIB is located, which was first noted by
\cite{Jenniskens+Desert1994} in the spectra of \object{HD\,30614}, \object{HD\,21389},
\object{HD\,190603}, and \object{HD\,183143}. This leaves only the \Ionw{He}{2}{4542} line
unaffected by DIB absorption.

\section{Radial velocity analysis}
\label{sect:RVs}

\begin{figure*}[ht]
\centering
\includegraphics[width=\textwidth]{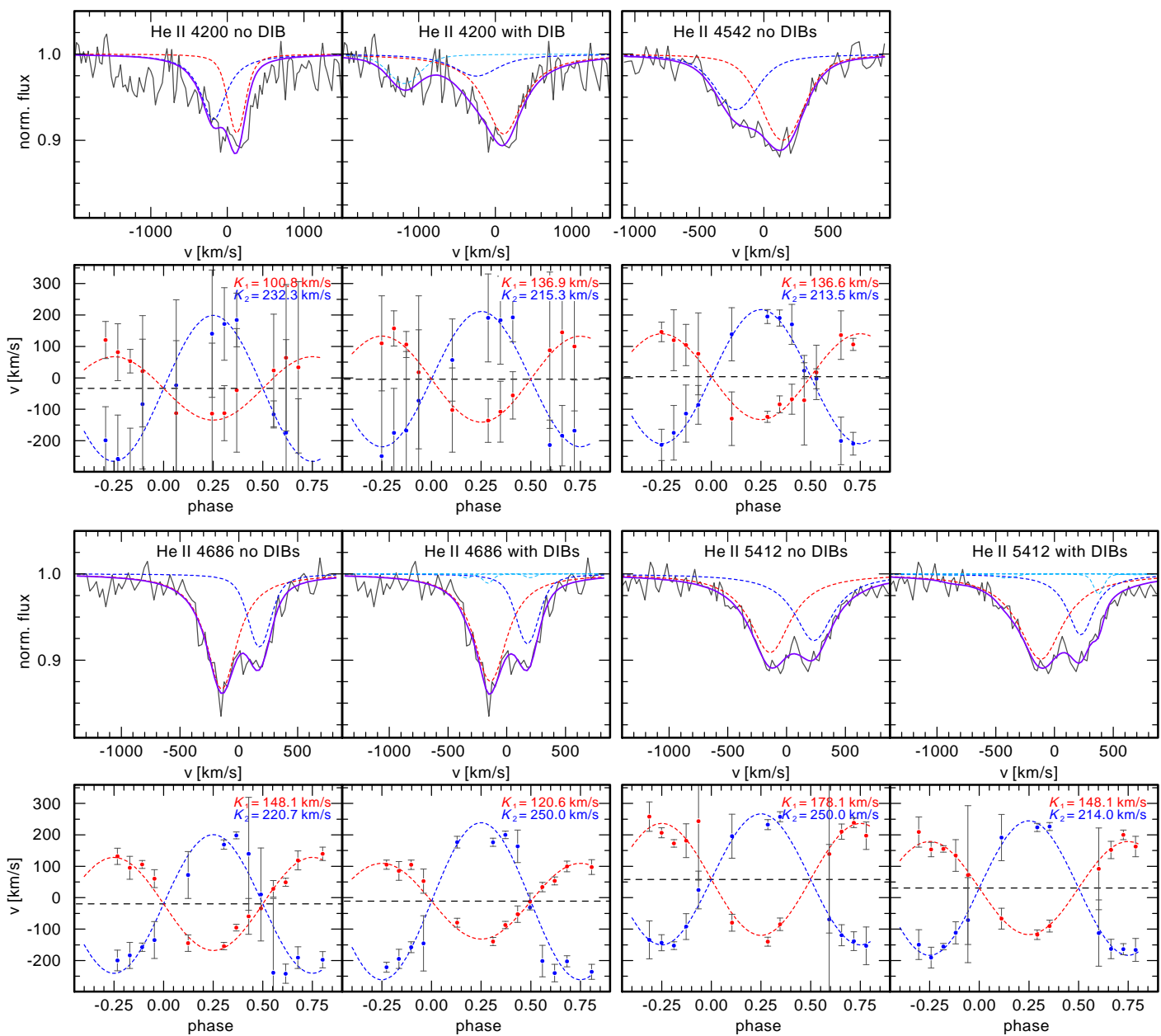}
\caption{Examples of Voigt profile fits (blue: secondary, red:
  primary, light blue: DIBs, purple: combined fit) to the four observed He\,{\sc ii}
  lines (gray). The top panels show observation \#2 taken around phase
  $-0.25$, the bottom panels show observation \#7 taken around phase
  $0.28$. RV curves (blue: secondary, red: primary) obtained for the
  respective He\,{\sc ii} lines with the RV amplitudes are shown below. The black, dashed line in RV curve plots indicates
the system velocity.}
\label{fig:rvs}
\end{figure*}

\subsection{Wavelength calibration accuracy of the OSIRIS spectra}
\label{sect:wavcal}
The accuracy of the wavelength calibration is a crucial point when
determining the RVs of a binary system. We used the nebular lines to check
the wavelength calibration of the OSIRIS spectra, as the RVs of these
lines should not change over the orbital period and correspond to the system velocity.
For that we first measured the RVs of the Balmer emission lines by fitting
them with a set of mathematical functions (Gaussians, Lorentzians, and polynomials) 
using SPAS (Spectrum Plotting and Analysing Suite, \citealt{Hirsch2009}).
The error determination is done by using the bootstrapping method.
We find that the line to line variations are small within a single exposure, suggesting an accuracy of the
wavelength calibration of 5\,km/s in the wavelength range $4101-4681$\AA. However,
the RVs of the Balmer lines from the different exposures show large
variations, indicating zero-point shifts of the wavelength calibration up to
54\,km/s (Table~\ref{tab:obs}).
Consequently, we corrected each observation for the RV measured by fitting
all three Balmer emission lines simultaneously, which leaves us with an artifical
system velocity of 0\,km/s.\\
It is worth mentioning that wavelength calibration exposures for the OSIRIS spectra were taken only
in the beginning of the observing run. Therefore, the 
velocity measured simultaneously from H\,$\delta$, H\,$\gamma$, and H\,$\beta$ from
observation \#1 ($66\pm1$\,km/s) should reflect the true system velocity. This value also
agrees with the system velocity of $70\pm8$\,km/s reported by \cite{Rodriguezetal2001}.\\
To check the wavelength calibration in the red part of the spectra
(i.e., around \Ionw{He}{2}{5412}, used by SG+15 to determine the masses), we
measured the RVs of the \Ionw{Fe}{2}{5376}
nebular line (closest nebular line to \Ionw{He}{2}{5412}) in the
zero-point corrected observations. We found variations up to 68\,km/s in the
different spectra. Since those lines are relatively weak, we could not
detect them in three of the fifteen observations and we also note that the average
uncertainty on the measured RVs of \Ionw{Fe}{2}{5376} are 20\,km/s. 
Therefore, we refrain from applying additional corrections to the spectra and merely
state that the wavelength calibration accuracy seems to get worse than
5\,km/s in the red part of the spectrum.

\subsection{Radial velocity amplitudes}
\label{sect:Ks}

Since Gaussian line profiles as used by SG+15 only provide a good
fit to the line cores of the \ion{He}{II} lines but not to their wings, we used Voigt profiles to
measure the RVs for both components. Using Python, Voigt profiles were
calculated via the Faddeeva function and fitted to the zero-point corrected
OSIRIS spectra using the non-linear least squares method of Levenberg-Marquardt
\citep{SciPy2001}. 
The semi-amplitudes of the RV curves ($K_1$, $K_2$) were then obtained by sinusoidal fitting of the 
individual RV measurements obtained for both components of the binary
system. The system velocity, $\gamma$, the orbital period, $P$, and the zero
point of the RV curve (the latter two within the uncertainties determined from
the light curves by SG+15) were allowed to vary, but required to
be the same for both stars.\\
Examples of the Voigt profile fits to the four observed He\,{\sc ii} lines
(gray lines) are shown in Fig.~\ref{fig:rvs}, along with the resulting RV
curves for each line. The red and blue lines correspond to the absorption lines and RV
curves of the primary and secondary, respectively. The purple line indicates
the combined fit. The black, dashed line in the RV curve plots indicates
the system velocity (remember we applied an artifical
system velocity of 0\,km/s based on the H\,{\sc i} nebular emission lines,
see \se{sect:wavcal}).\\
First, the RV fitting was performed assuming only two Voigt profiles for each
\ion{He}{II} feature corresponding to the absorption lines of the two stars. In
case of \Ionw{He}{2}{4686} and \Ionw{He}{2}{5412} we next included
additional, fixed Voigt profiles in order to simulate the DIBs (light blue,
dashed lines in Fig.~\ref{fig:rvs}). The equivalent widths and full widths at half maximum of
these DIBs were required to be the same as reported by
\cite{Hobbs+2008} for HD\,204827. The DIB blue-ward of \Ionw{He}{2}{4200}, which also
blends with this line, is clearly visible in the co-added OSIRIS spectrum
(Fig.~\ref{fig:rvs}). Therefore, we obtained the Voigt profile for this DIB
directly from the co-added OSIRIS spectrum, and used this line profile in each
subsequent RV fit.\\
Neglecting DIBs, we find for \Ionw{He}{2}{5412} similar RV
amplitudes of $K_1=178\pm17$\,km/s and $K_2=209\pm18$\,km/s, which is
close to the values derived by SG$+$15 by Gaussian fitting of the 
\Ionw{He}{2}{5412} absorption lines ($K_1=206\pm8$\,km/s and
$K_2=206\pm12$\,km/s). We note that we obtain a system velocity much
larger than zero ($\gamma=58\pm9$\,km/s), indicating already a problematic
result. The picture, however, changes noticeably if additional
DIB absorption lines are included in the RV fitting process. We then obtain
very distinct RV amplitudes of $K_1=148\pm9$\,km/s and $K_2=214\pm10$\,km/s,
i.e. we find that the RV amplitude of the primary star could be $30$\,km/s
smaller. The value for the system velocity ($\gamma=31\pm5$\,km/s) improves,
but is still clearly larger than zero. This likely indicates that DIBs in \hen
are different to HD\,204827 and/or that the wavelength calibration in the red
part of the spectrum becomes slightly worse (see \se{sect:wavcal}).\\
For \Ionw{He}{2}{4686} we find two different RV amplitudes
($K_1=148\pm22$\,km/s and $K_2=221\pm24$\,km/s) even 
if we neglect the DIBs. Including DIBs in the fitting, the differences become
even more noticable ($K_1=121\pm18$\,km/s and $K_2=250\pm21$\,km/s). Also in
this case, the RV curve fitting suggests system velocities which are smaller than
zero ($\gamma=-20\pm10$\,km/s, when no DIBs are considered, and
$\gamma=-11\pm9$\,km/s, when the DIBs are included). However, the deviation
from zero is not as drastic as in the case of \Ionw{He}{2}{5412}.\\
An interesting point to notice is that the line profiles of 
\Ionw{He}{2}{5412} of both the primary and secondary are very similar if DIBs are
neglected. Including DIBs in the fits, the line of the secondary becomes
much weaker (see the right hand panel in the second to last row of
Fig.~\ref{fig:rvs}). For \Ionw{He}{2}{4686} (which is blended with weaker
DIBs than \Ionw{He}{2}{5412}) and \Ionw{He}{2}{4542} (not blended with any
DIB) it is already evident from the observed line profiles, that the line of
the secondary must be weaker than the line of the primary.\\
The RV amplitudes of \Ionw{He}{2}{4542} are of greatest interest as it is the
only line not blended with any DIB. For this line we obtain again very
different RV amplitudes of $K_1=137\pm12$\,km/s for the primary and
$K_2=214\pm14$\,km/s for the secondary. This supports the results from
the RV fits of \Ionw{He}{2}{4686} and \Ionw{He}{2}{5412} if DIBs are
included. We also stress that in the case of \Ionw{He}{2}{4542}, we obtain a system
velocity of only $4\pm6$\,km/s, consistent with zero.\\
The blue parts of the OSIRIS spectra covering \Ionw{He}{2}{4200} have a
lower S/N, resulting in larger uncertainties of the individual RV
measurements. If we neglect the absorption of the broad DIB blue-ward of
\Ionw{He}{2}{4200}, we obtain RV amplitudes of $K_1=101\pm25$\,km/s and
$K_2=232\pm34$\,km/s, and $\gamma=-34\pm10$\,km/s. If we, however, include our DIB model which we 
obtained directly from the co-added spectrum (see above), we
end up with RV amplitudes of $K_1=137\pm18$\,km/s and $K_2=215\pm21$\,km/s,
confirming the results from \Ionw{He}{2}{4542} surprisingly well. Also in
this case we find that the system velocity is very small ($-4\pm9$\,km/s) and
consistent with zero.\\
In summary, if DIBs are not included in the RV fitting, we end up with
conflicting RV semi-amplitudes for the four \ion{He}{II} lines. However, when including the
DIBs we obtain consistent results. Since our DIB models for \Ionw{He}{2}{4686}
and \Ionw{He}{2}{5412} may not be perfect assumptions (as indicated from the 
non-zero system velocities), the RV amplitudes derived from \Ionw{He}{2}{4200} and
\Ionw{He}{2}{4542} should be the ones to rely on. For these lines very
distinct RV amplitudes are found as opposed to the findings of SG+15.

\section{Atmospheric analysis}
\label{sect:atmo}

For the spectral analysis we restricted ourselves to the OSIRIS observation
\#2. This is because this observation was taken closest to maximum RV separation
and, hence, smearing of the lines due to the orbital motion (i.e., the change of the RV over
the duration of the exposure) is only a few km/s (close to phase 0 and 0.5 the
orbital smearing reaches about 78\,km/s). Also in observations \#2 none of the
four He\,{\sc ii} lines are contaminated with emission lines.
The spectrum was decomposed by subtracting the line profiles obtained from
the RV fitting (\se{sect:RVs}) for the DIBs and the other star
from the observation.\\
For the model calculations we employed the T{\"u}bingen non-LTE 
model-atmosphere package
(TMAP\footnote{http://astro.uni-tuebingen.de/\textasciitilde TMAP}, 
\citealt{werneretal2003, rauchdeetjen2003, Tmap2012}) which allows plane-parallel, non-LTE, fully 
metal-line blanketed model atmospheres in radiative and hydrostatic equilibrium to be computed. Model atoms were 
taken from the T{\"u}bingen model atom database
TMAD\footnote{http://astro.uni-tuebingen.de/\textasciitilde TMAD}.
Metal-free model grids were calculated for six different He abundances ($\log\mathrm{He/H}=+2, +1, 0, -1, -2$, and $-3$,
logarithmic number ratios). Each grid spans from $T_{\rm  eff}=30\,000-70\,000\,{\rm K}$ (step size 2500\,K)
and from $\log{g}=3.75-6.0$ (step size 0.25\,dex). Models above the Eddington
limit (i.e., $T_{\rm eff}>60\,000$\,K for \loggw{4.25}, $T_{\rm  eff}>50\,000$\,K for
\loggw{4.00}, and $T_{\rm  eff}>47\,500$\,K for \loggw{3.75}) were not calculated.
To calculate synthetic line profiles, we used Stark line-broadening tables provided by \cite{Barnard1969} 
for \Ionww{He}{1}{4026, 4388, 4471, 4921}, \cite{Barnard1974} for \Ionw{He}{1}{4471}, and \cite{Griem1974} 
for all other \ion{He}{I} lines. For \ion{He}{II}, we used the tables provided
by \cite{Schoening1989}, and for H\,{\sc i} tables provided by
\cite{TremblayBergeron2009}. For He\,{\sc ii} 20 levels were considered in non-LTE,
for He\,{\sc i} 29 levels, and for H\,{\sc i} 15 levels.\\
To derive effective temperatures, surface gravities, and He abundances we
fitted simultaneously all four decomposed He\,{\sc ii} lines of each star.
The parameter fit was performed by means of a $\chi^2$ minimization technique with SPAS (Spectrum 
Plotting and Analysing Suite, \citealt{Hirsch2009}), which is based on the
FITSB2 routine \citep{Napiwotzki1999}. Although we do not expect the system to
be fully synchronized shortly after a common envelope event, both stars
have likely high rotational velocities. Therefore, we considered the projected
rotational velocity $v \sin i$ as a forth parameter in our fit.\\
Our initial fit assumes a flux ratio of one, and the effective temperatures,
surface gravities, He abundances, and rotational velocities were considered
as free parameters. Based on these results we performed light curve fits (\se{sect:photo}) and
used the updated flux ratios and rotational velocities (assuming a synchronized
system) to repeat the spectral fits. This iterative process was repeated
until a good agreement between the results from the light curve fitting and
the spectral analysis was obtained. In our final spectroscopic fit we assume a
flux ratio of 1.4 and projected rotational velocities of $v \sin i= 156$\,km/s for the
primary and $v \sin i= 133$\,km/s for the secondary.

\begin{figure}[t]
\centering
\includegraphics[width=\columnwidth]{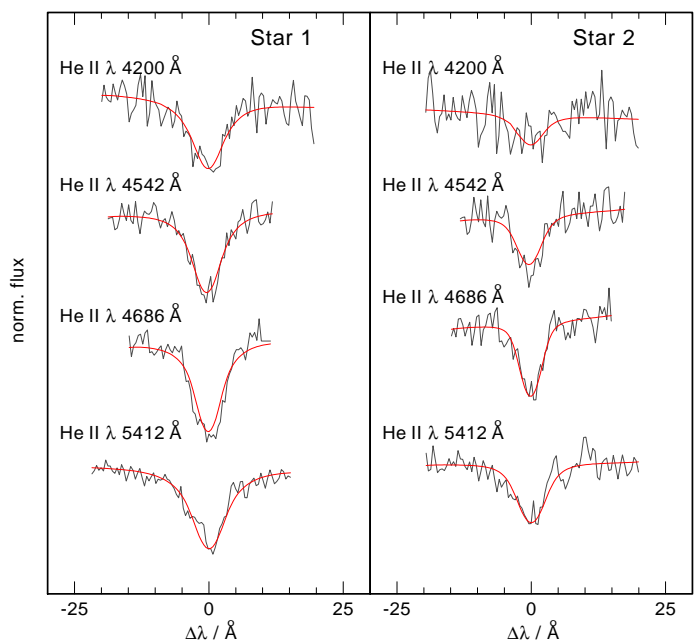}
\caption{He\,{\sc ii} lines of the decomposed OSIRIS spectrum \#2 (gray) shown along
  with the best the fit TMAP models (red).}
\label{fig:FIT}
\end{figure}

Our best fits to spectrum \#2 are shown in Fig.~\ref{fig:FIT} and the results of our analysis
are summarized in Table~\ref{tab:parameters}. We note that our best fit also
reproduces very well the wings of the Balmer and \ion{He}{I} lines (Fig.~\ref{fig:Osiris2bestfit}).
The effective temperatures (\Teff$=39555$\,K and \Teff$=40858$\,K, for the primary and
secondary, respectively) and surface gravities (\loggw{4.50} and \loggw{4.62}) obtained for both stars are
found to be very similar and agree well within the error limits with the
results from the light curve fitting (see \se{sect:photo} and Table~\ref{tab:parameters}).
The He abundance of the primary ($X_{\mathrm{He}}=-0.16\pm0.10$, logarithmic mass fraction), which has
the stronger lines, is found to be super solar
($X_{\mathrm{He}_{\odot}}=-0.60$, \citealt{Asplundetal2009}), while the secondary
has a slightly sub-solar He abundance ($X_{\mathrm{He}}=-1.01\pm0.20$).\\ 
Our effective temperatures are larger than the ones reported by SG+15, who
derived the effective temperatures from light curve fitting. The narrow
temperature range ($30-40$\,kK) adopted by SG+15, however, is not valid as already
pointed out by \citet{GB+2016}. SG+15 established the upper limit of $40$\,kK based on the
absence of \ion{He}{II} emission lines, but there are many CSPNe with even higher \Teff\
and which also lack \ion{He}{II} nebular lines.
The difference to our previously reported values for the atmospheric parameters
\citep{Finch+2018, Reindl+2018, Finch+2019} is a consequence of the
rotational velocity which was neglected in our previous fits, as well as the
extended model grid, the avoidance of observations which are noticeably
affected by smearing of the lines due to the orbital motion of the system,
the updated flux ratio of the system revealed by the light curve analysis,
and the consideration of the DIB absorptions.\\
We emphasize that an accurate spectral analysis of the system is very
challenging. This is because we lack the knowledge of the exact rotational
velocities (the intrinsically broad \ion{He}{II} lines are not a good approach to
determine the rotational velocity, especially if only medium-resolution spectra are
available), neglect the special geometry of the system, the incoming radiation of the
other star, as well as metal opacities in our model
atmosphere calculations. Finally the exact equivalent widths of the DIBs blending with
\Ionw{He}{2}{4686} and \Ionw{He}{2}{5412} are not known, adding another
uncertainty. The errors given in Table~\ref{tab:parameters} therefore not
only include the formal fitting errors, but also estimates on the systematic
uncertainties mentioned above.

\section{Light curve modelling}
\label{sect:photo}

\begin{figure}[t]
\centering
\includegraphics[width=\columnwidth]{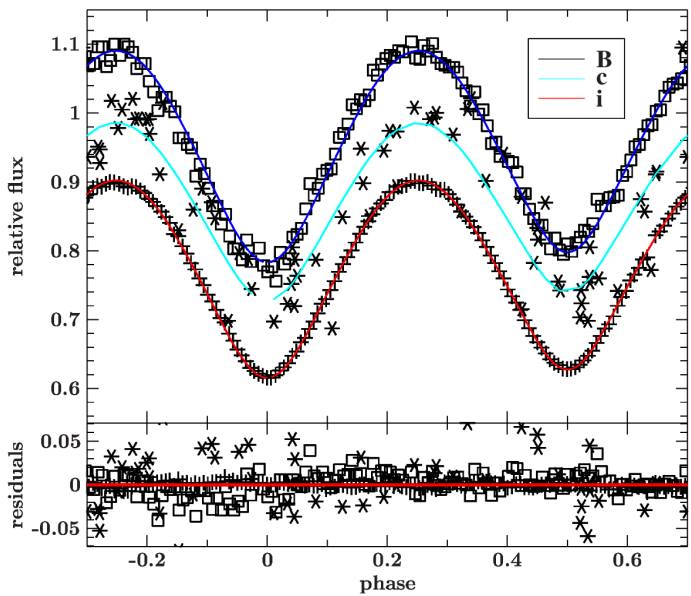}
\caption{Relative fluxes of the Johnson B-band (squares), ATLAS c-band (asterisks), and Sloan i-band
  (crosses) light curves compared  to our best fit MORO models (blue,
  light-blue, and red, respectively). The light curves are shifted vertically
  for clarity. Residuals are shown at the bottom.}
\label{fig:lightcurve}
\end{figure}

The analysis of the light curves was carried out simultaneously in the Johnson
B-band, Sloan i-band, and ATLAS c-band filters. Because of its poor S/N, the
ATLAS o-band light curve was omitted from our analysis. 
First fits of the light curves showed that the mass ratio is not constrained
by the shape of the light curves. This is due to the significant degeneracies
of the many dependent parameters used in the light curve analysis, which
permits in many cases the determination of the mass ratio by light curve
analysis \citep[e.g.,][]{Schaffenroth+2014}. Only when ellipsoidal
deformation is visible can the mass ratio be constrained \citep[e.g.,][]{Kupfer+2017}.
Therefore, we fixed the mass ratio of the system to the one which was
derived by the RV analysis of \Ionw{He}{2}{4542} and used the effective
temperatures derived by the spectral analysis as starting values.\\
For the analysis we used MORO (Modified Roche Program, see
\citealt{Drechsel1995}). It is based on the Wilson-Devinney mode 3 code,
that is used for overcontact systems \citep[see][]{KallrathEugene2009} using a
modified Roche model considering the influence of the radiation pressure on the 
shape of the stars. The program assumes equal Roche potentials, limb darkening
and gravitational darkening coefficients for both stars. The optimization of
parameters is achieved by the simplex algorithm. The gravitational darkening
parameters were fixed at 1.0 as predicted for
radiative envelopes \citep{vonZeipel1924}. The limb darkening coefficients were
taken from \cite{ClaretBloemen2011} using the value closest to the parameters determined
by the spectroscopic analysis for the different filters respectively. As both
stars have comparable temperatures we also fixed the albedo to 1.0.
We also considered a third light source, accounting for the nebular continuum
and line emission. By varying the radiation pressure parameter,
inclination, temperatures, Roche potentials, luminosity ratio of both stars,
and the third light contribution, the curves are reproduced nicely. We note that our fit reproduces
the light curves better than the one of \cite{GB+2016}, and also slightly
better than the model of SG+15.\\
Our best fits to the light curves are shown in Fig.~\ref{fig:lightcurve} and
the results of our analysis are summarized in Table~\ref{tab:parameters} (see
also Table~\ref{tab:light_full} for all parameters of the best light curve fit).
We find a relative luminosity $\frac{L_1}{L_1+L_2}$ of 58.37\% in the B-band,
similar effective temperatures for both stars (\Teff$\approx40\,kK$), and that
the mean radius of the secondary is 15\% smaller than the radius of the
primary. We derived an additional constant flux component of 0.8\% in
B, 2.8\% in i, and 20.5\% in the ATLAS c-band\footnote{We note, that for compact
nebulae, the flux contribution of the nebular lines can
be significant when broad band filters are used (\citealt{ShawKaler1985, GathierPottasch1988}). 
For example the V-band magnitude of the CSPN of the Stingray Nebula measured with the
Hubble Space Telescope and, thus, resolving the CSPN, is four orders of
magnitude smaller than what is measured from the ground
\citep{Bobrowsky1998, Schaefer+2015}. Therefore, the high additional flux
contributions in the ATLAS c-band, which covers numerous nebular lines
(e.g., [\ion{O}{III}] $\lambda\,5007$\,\AA), is not surprising.}. 
Combining it with the results from the RV curves the absolute parameters $M,
R$ could be derived (Table~\ref{tab:parameters}, see also
\se{sect:masses}). The errors were determined by a bootstraping method
\citep[see][]{Schaffenroth+2014} and represent only the statistical error resulting
from the noise in the light curves and do not consider the degeneracies in the
light curve.

\begin{figure}[t]
\centering
\includegraphics[width=\columnwidth]{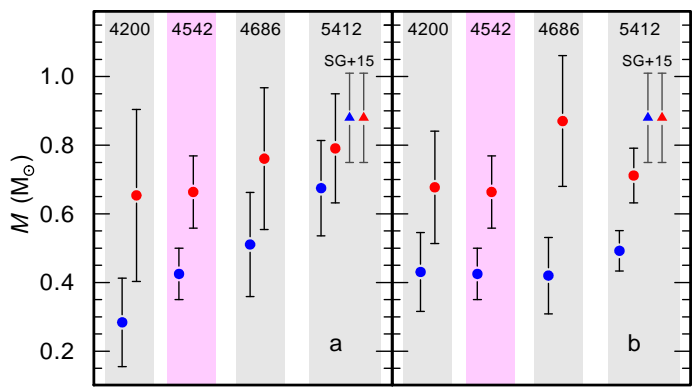}
\caption{Dynamical masses of the primary (red dots) and the secondary (blue dots) as
  determined with the inclination angle from our light curve analysis and 
  our RV fitting of the four He\,{\sc ii} lines excluding (panel a) and including
  DIBs (panel b). Masses obtained from He\,{\sc ii} lines that blend with DIBs are highlighted in
  grey, He\,{\sc ii}\,$\lambda$\,4542\,\AA\ which is not blended with any DIB is
  highlighted in pink. The blue and red triangles are the masses reported by SG$+$15.}
\label{fig:masses}
\end{figure}

\section{Dynamical masses and evolutionary status}
\label{sect:masses}
Since the inclination of the system ($i=63.59\pm0.54$\,\textdegree) can be constrained well
from the light curve fitting, the dynamical masses of the two stars can be
calculated via the binary mass function
$$f(M_1, M_2)=\frac{K^3_{1}\,P}{2\pi G}=\frac{M_2 \sin^3i}{(1+\frac{M_1}{M_2})^2}.$$
Our results are shown in Fig.~\ref{fig:masses}. In panel a, we
show the dynamical masses as obtained using our Voigt profile fitting routine and neglecting DIBs. It can be
seen, that our masses obtained from \Ionw{He}{2}{5412} with the Voigt
profile RV fitting and after applying the zero point correction
($M_1=0.79\pm0.16$\,\Msol\ and $M_2=0.67\pm0.14$\,\Msol) agree within the error limits
with the results of SG+15, who find $0.88\pm0.13$\,\Msol\ for both stars. However, for all other lines 
contradictory results are found.\\
When DIBs are included in the RV fitting significantly different results are
obtained for the masses of the two CSPNe (e.g., $M_1=0.71\pm0.08$\,\Msol\ and
$M_2=0.49\pm0.06$\,\Msol for \Ionw{He}{2}{5412}, panel b in Fig.~\ref{fig:masses}).
In this case we find that the masses from the different \Ion{He}{2} lines agree with each
other, but no longer with the results from SG+15. This is a consequence of the
zeropoint corrections, using Voigt profiles instead of Gaussians in the RV
fitting, and most importantly the inclusion of DIBs when determining the
RVs.\\
We stress that masses obtained from \Ionww{He}{2}{4200, 4542}
are the ones to be trusted. This is because only \Ionw{He}{2}{4542} is not blended with any
DIB and for \Ionw{He}{2}{4200} a good fit to the DIB which blends with this
line can be found, though the S/N in this part of the spectrum is rather
poor. For \Ionw{He}{2}{4686} and \Ionw{He}{2}{5412} we can only
assume the DIBs are of about the same strength as in HD\,204827 based on the
similar reddening. However, the equivalent widths of the DIBs blending with these lines might
be different, for example because of a different chemical composition of the
interstellar or circumstellar medium.\\
The masses obtained for \Ionw{He}{2}{4542} are
$M_1=0.66\pm0.11$\,\Msol\ and $M_2=0.42\pm0.07$\,\Msol, and agree very well
with the masses obtained from \Ionw{He}{2}{4200}
($M_1=0.68\pm0.16$\,\Msol\ and $M_2=0.43\pm0.11$\,\Msol).
This is a striking result, as with the masses derived from \Ionw{He}{2}{4542}
the total mass of the system ($M=1.08\pm0.18$), no longer exceeds the 
Chandrasekhar mass limit.\\
The total mass of the system is still high enough that a merger of the system
will occur within a Hubble time. However, with the combined dynamical mass of the system
no longer exceeding the Chandrasekhar mass limit, the merger will not
produce a SN\,Ia via the traditional double-degenerate channel \citep{HanPodsiadlowski2004}.
The individual masses of the two CSPNe are also too small for a reasonable production of 56Ni (which
determines the explosion brightness) in case of a dynamical explosion during
the merger process \citep{Pakmor+2013, Shen+2018}. Thus, the merging event of Hen\,2-428 will not be
identified as a SN Ia.\\
Most likely, the merger of \hen will then lead to the formation of a
(He-rich) RCB star $\to$ EHe star $\to$ massive O(He) star $\to$
CO white dwarf \citep{Schwab2019, Shen2015, Zhang2014}. If both stars should have
CO-cores at the time of the merger, the formation of a star with a C/O-dominated
atmosphere could be possible. This would make \hen a promising progenitor for
the CO-dominated hot white dwarf stars H1504+65 and RXJ0439.8$-$6809 \citep{WernerRauch2015},
and for the C-dominated hot DQ white dwarfs \citep{Kawka+2020}. 
The formation of an AM CVn type system via the double white dwarf channel
\citep{Paczynski1967}, that will end up in a faint
thermonuclear supernova, however, seems very unlikely due to the high mass
ratio of the system \citep{Nelemans+2001, Marshetal2004}.

\begin{figure}[t]
\centering
\includegraphics[width=\columnwidth]{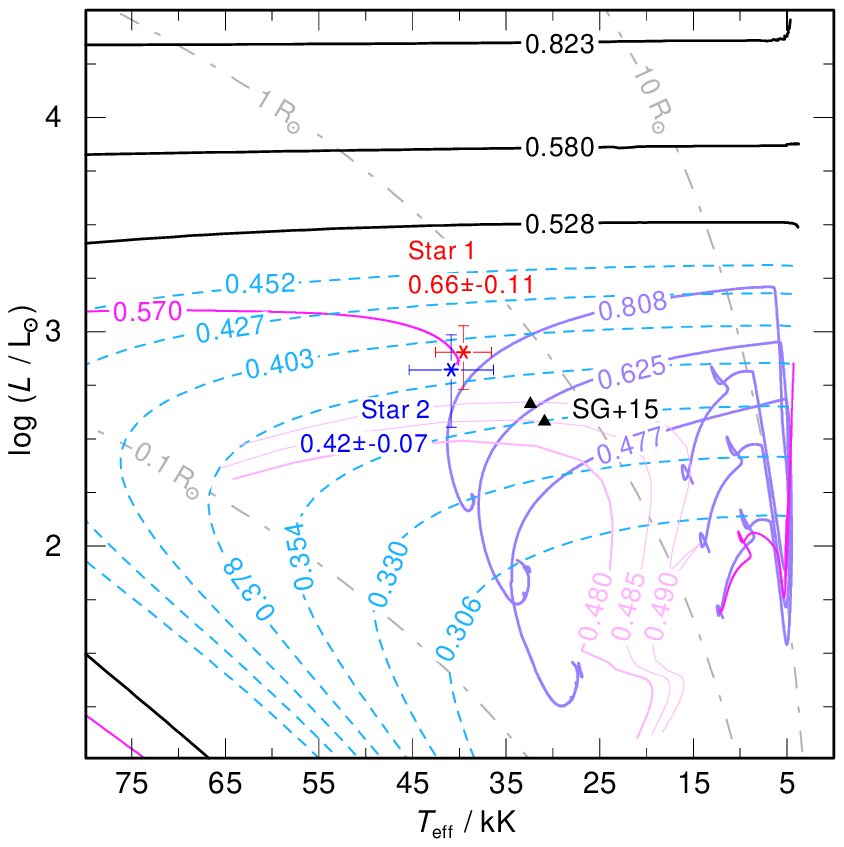}
\caption{Locations of the two CSPNe and their dynamical masses (primary is shown in red, the secondary in blue) in the HRD compared to stellar
  evolutionary tracks. Black lines indicate H-shell burning post-AGB tracks from
  \cite{MillerBertolami2016}, dashed, light-blue lines post-RGB tracks from
  \cite{Hall+2013}, purple lines stripped star evolutionary tracks from
  \cite{Goetberg+2018}, and in pink post-EHB evolutionary tracks from
  \cite{Dorman1993}. The magenta line indicates an evolutionary track of a
  He-shell burning stripped post-early AGB star. The gray dashed-dotted lines indicate radii of 0.1, 1, and 10\,$R_\odot$}
\label{fig:HRD}
\end{figure}

In Fig.~\ref{fig:HRD}, we show the locations of the two CSPNe in the
Hertzsprung Russell diagram (HRD), as derived with the effective temperatures
from our spectroscopic analysis and the radii from the light curve analysis
(primary is shown in red, the secondary in blue).
It can be seen, that the luminosities and radii of the two stars are too
low for what is expected for normal post-asymptotic giant branch (AGB) stars
(black lines indicate H-shell burning post-AGB tracks from \citealt{MillerBertolami2016}).
The dynamical mass and location in the HRD of the secondary agrees with
predications for post-RGB stars (light-blue lines, \citealt{Hall+2013}), while the dynamical
mass of the primary is too high for this scenario. The secondary could also be
a post-extreme horizontal branch (post-EHB) star (pink lines in
Fig.~\ref{fig:HRD} are post-EHB tracks from \citealt{Dorman1993}), while the 
the mass of the primary is again too high for this scenario. 
It is worthwhile mentioning that the mass of the remaining H
layer ($\le0.001$\Msol) of EHB stars is much too low to produce a nebula
at the end of the He-core burning stage. Thus, it is not possible for both
stars to be post-EHB stars.\\
The solid purple lines in Fig.~\ref{fig:HRD} correspond to evolutionary tracks for
stars stripped through Roche-lobe overflow and were calculated by
\cite{Goetberg+2018}. The stars had initial masses of 3.65, 2.99, and
2.44\,\Msol\ and the tracks show the evolution from central H-burning, the mass
transfer phase, consequent blue-ward evolution, until He-core burning is
reached. The mass of the secondary is too small in order to descend from such a
star, while the primary could be a candidate for being a stripped He-star,
shortly before the central He-core burning phase. The surface He abundances
predicted by \cite{Goetberg+2018} for this evolutionary stage
($X_{\mathrm{He}}\approx-0.15$) matches surprisingly well with what we find in our
spectroscopic analysis of the primary ($X_{\mathrm{He}}=-0.16$).\\
We stress that the comparison with these evolutionary tracks should
be treated with caution, as obviously none of these models can account for the
real evolution of \hen. Stable Roche-lobe overflow, for example, cannot account
for the short orbital period and over-contact nature of the system, meaning
the latest mass-transfer phase must have ended in a common envelope ejection.
In addition it is not clear to what extent the evolutionary tracks (and the
mass-radius relationship) are altered for over-contact systems. 
Short orbital period ($P<1$\,d) low-mass main sequence stars, for example,
show an inflation by 10\% \citep{Kraus+2011}, thus it could be possible that
the radii and luminosities of two CSPNe of \hen are also too large compared to
single-star evolutionary tracks. For the massive over-contact system \object{VFTS\,352}
(temperature-wise very similar to \hen), it was found that single-star models
predict effective temperatures which are 6\% lower than what would be expected
from the dynamical masses \citep{Almeida+2015}.\\

With all caveats in mind we can still make an educated guess of what
the evolution of the system might have been. In light of the actual
close configuration of the system and the presence of surrounding
material we know that the last mass transfer episode was unstable and
led to the formation and ejection of a common envelope. In addition,
the derived dynamical masses ($M_1=0.66\pm 0.11$\,\Msol\ and $M_2=0.42\pm0.07$\,\Msol), 
temperatures and luminosities (and radii), together with a 
comparison with stellar evolution models suggests that the secondary
(in the following \mbox{Star\,2}) has a post-RGB like structure, meaning it
has a degenerate He-core surrounded by a H-burning shell and a thin envelope
on top. The nature of the more massive component (in the following
\mbox{Star\,1}) is less certain. The actual mass of the object, however,
indicates that before the last mass transfer episode \mbox{Star\,1} was not
a low-mass RGB star and its mass before the last mass transfer
episode was beyond that needed for non-degenerate He ignition. As
\mbox{Star\,2} is already a low-mass evolved star we can safely conclude
that \mbox{Star\,1} was originally the less massive component and increased
its mass during a previous (first) mass transfer episode. This implies
that the total initial mass of the system was necessarily
$M_1^i+M_2^i \lesssim 3.6$\,\Msol\ \citep{Bressan+2012}, and $M_2\lesssim 3.2$\,\Msol\ 
before the common envelope episode.\\
As mentioned above, the inferred surface properties and mass of
\mbox{Star\,1} are in good agreement with the predicted evolution of a
intermediate mass star that was stripped of in its post-main sequence evolution
before the ignition of He-core burning. A serious shortcoming of this
scenario is that, for this to happen, \mbox{Star\,1} needs to fill its Roche
lobe before He-ignition, but intermediate-mass stars with $M_2\approx 3.2$\,\Msol\
reach at most $R_2\sim 45$\,\Rsol\ before He-ignition. With
a $q$-value of $q\approx 3.2$\,\Msol$/0.4$\,\Msol$\,=8$ that means that the
Roche lobe of the 0.4\,\Msol\ companion should have been $R_1\lesssim 18$\,\Rsol\
at the end of the first (stable) mass transfer episode. Due
to the tight core mass-radius relation of RGB stars, and the fact that
during Roche lobe overflow $R_\star\simeq R_{\rm Roche}$ \citep{Han+2000},
\mbox{Star\,2} would have probably been peeled off well before the
mass of the degenerate He-core reached $\approx 0.3 M_\odot$. Given that
$M_2=0.42\pm 0.07\,M_\odot$, this scenario seems unlikely.

\begin{figure}[t]
\centering
\includegraphics[width=\columnwidth]{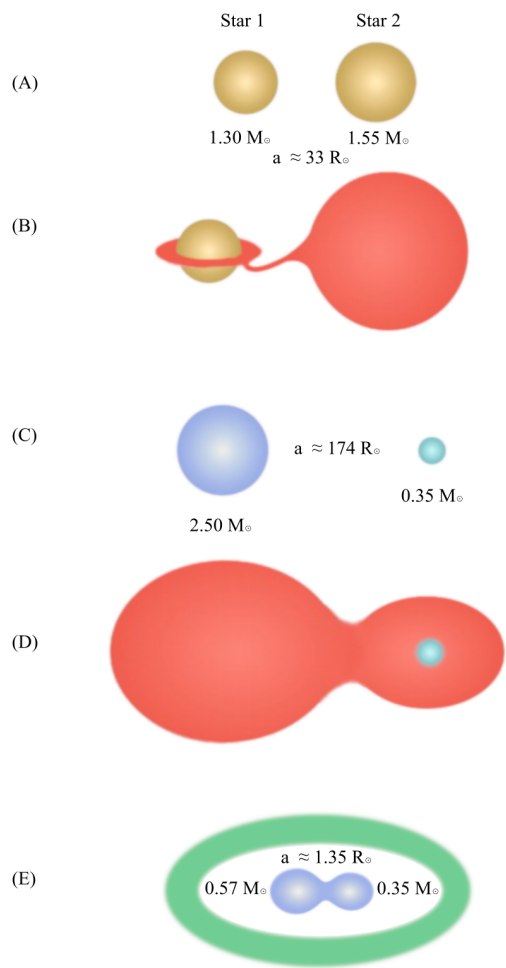}
\caption{Sketch of a possible evolutionary scenario for \hen.}
\label{fig:scenario}
\end{figure}

Interestingly, stars with masses in the range 2.5 to 3\,\Msol\
expand in the early-AGB phase\footnote{This is, after the end of
core-helium burning and before the development of thermal pulses.} to
very large radii of $R_\star>100$\,\Rsol. Moreover, during the early
AGB, a star in this mass range shuts down its H-burning shell, making the
stripping of the H-rich material easier, which in the light of the
high He enrichment found in the primary of the system makes this
scenario more compeling. The magenta line in Fig.~\ref{fig:HRD} shows the evolution of a 0.57\,\Msol\
post-early AGB model constructed by artificially stripping
the envelope of a 2.5\,\Msol\ star once it reached the luminosity of
\mbox{Star\,1}. The surface He mass fraction of the model is $X_{\rm He}\approx -0.28$.\\
Figure~\ref{fig:scenario} shows a toy model for such scenario. Lets assume that we start
the evolution with a pair of low-mass stars in a relatively close
orbit (panel A in Fig.~\ref{fig:scenario}, $M_2^i=1.55$\,\Msol,
$M_1^i=1.3$\,\Msol, $a \simeq 33$\,\Rsol). As soon as the more massive star ends its main
sequence evolution it will evolve into the RGB and when it reaches
$R_1\simeq 13$\,\Rsol\ it will start to transfer mass to its companion
(Panel B in Fig.~\ref{fig:scenario}). Due to the low mass ratio of the system at that
point ($q \lesssim 1.19$) mass transfer will be stable, and as soon as
$M_1<M_2$ it will evolve on a nuclear timescale \citep{Podsiadlowski2014},
and stable mass transfer continues as \mbox{Star\,2} evolves on the RGB. If
the envelope of \mbox{Star\,2} is removed once its He-degenerate core
reaches $0.35\,M_\odot$ the star will contract and form a $M^f_2\simeq 0.35$\,\Msol\
He-core white dwarf. Under the simplifying assumption
that mass loss is conservative \citep{PostnovYungelson2014} our system
would be composed of a $0.35$\,\Msol\ He-core white dwarf, and a 
rejuvenated $M_1^{\rm rj}=2.5$\,\Msol\ main-sequence companion,
separated by $a\simeq 174$\,\Rsol. The Roche lobe of \mbox{Star\,1} under
such situation would be of $R^1_{\rm Roche}\simeq 95$\,\Rsol\footnote{We note however that this is just a toy model, as the
sequence in Fig.~\ref{fig:HRD} was stripped on the AGB at $R\simeq 50$\,\Rsol\ in
order to match the luminosity of the primary component.}. \mbox{Star\,1}
will then end its main sequence phase, and go to the He-core burning phase
without interacting with its companion. But once He-core burning is
finished, the star will evolve to the early AGB. In isolation a
2.5\,\Msol\ star would expand to about $R^1\approx 170$\,\Rsol\ before
developing thermal pulses, but due to the presence of its companion as
soon as \mbox{Star\,1} fills its Roche lobe at $R^1_{\rm Roche}\simeq 95$\,\Rsol\
it will start transferring mass. Given the extreme mass ratio of the
system ($q=M_1^{\rm rj}/M^f_2 \simeq 7.14$) mass transfer will be
highly unstable, leading to the formation of a common envelope (Panel
D in Fig.~\ref{fig:scenario}), the shrinking of the orbits and the final ejection of
the common envelope.  The current state of the system would be an
overcontact close binary system composed of the post-early AGB core of
$M^f_1\simeq 0.57$\,\Msol, a post-RGB core of $M^f_2\simeq 0.35$\,\Msol\ with
its envelope reheated by the last mass transfer episode, and a
surrounding PN composed of the ejected material (Panel E in Fig.~\ref{fig:scenario}).

\begin{table}[t]
\caption{Orbital and stellar parameters of \hen. He abundances are given in logarithmic mass fractions.}
\begin{tabular}{l r@{\,$\pm$\,} l r@{\,$\pm$\,} l} 
\hline\hline
\noalign{\smallskip}
                 & \multicolumn{2}{c}{Primary}   & \multicolumn{2}{c}{Secondary}      \\
\noalign{\smallskip}
\hline
\noalign{\smallskip}
$P$ [days]$^{(a)}$       &  \multicolumn{4}{c}{0.1758\,$\pm$\,0.0005}\\
\noalign{\smallskip}
$\gamma$ [km/s]       &  \multicolumn{4}{c}{66\,$\pm$\,1}\\
\noalign{\smallskip}
$q \equiv M_2/M_1$     &  \multicolumn{4}{c}{$0.64^{+0.25}_{-0.18}$}\\
\noalign{\smallskip}
$i $ [\textdegree]     &  \multicolumn{4}{c}{63.59\,$\pm$\,0.54}\\
\noalign{\smallskip}
$a $ [\Rsol]           &  \multicolumn{4}{c}{1.35\,$\pm$\,0.07}\\
\noalign{\smallskip}
$T_{\rm eff}$ [K] (Spec.) &  $39555$&$3000$ & $40858$&$4500$\\
\noalign{\smallskip}
$T_{\rm eff}$ [K] (LC)    &  $40179$&$370$ & $40356$&$175$\\
\noalign{\smallskip}
$\log{g}$  (Spec.)     &  $4.50$&$0.30$  &  $4.62$&$0.30$ \\
\noalign{\smallskip}
$\log{g}$  (LC)        &  $4.69$&$0.03$  &  $4.64$&$0.04$ \\
\noalign{\smallskip}
$X_{\mathrm{He}}$        &  $-0.16$&$0.10$  & $-1.01$&$0.20$ \\
\noalign{\smallskip}
$v_{\rm rot}$ [km/s]$^{(b)}$    &  \multicolumn{2}{c}{$174$} & \multicolumn{2}{c}{$148$}\\
\noalign{\smallskip}
$K_{4542}$ [km/s]       &    $136.6$&$12.0$  & $213.5$&$13.7$\\
\noalign{\smallskip}
$M_{4542}$ [\Msol]      &  $0.66$&$0.11$  & $0.42$&$0.07$ \\
\noalign{\smallskip}
$R$ [mean, \Rsol]      &  $0.603$&$0.038$  & $0.514$&$0.033$ \\
\noalign{\smallskip}
$L$ [\Lsol]            &  $803$&$264$       & $665$&$305$ \\
\noalign{\smallskip}
\hline
\end{tabular}
\tablefoot{~\\
\tablefoottext{a}{Taken from SG+15.}
\tablefoottext{b}{Assuming a synchronized system.}
}
\label{tab:parameters}
\end{table}

\section{Summary and conclusion}
\label{sect:discussion}

We performed a detailed reanalysis of the alleged type Ia supernova
progenitor \hen. Our study reveals that the red-excess
reported by \cite{Rodriguezetal2001} is merely a consequence of the
\cite{Cardelli+1989} reddening law used in their work. Fitting
the IDS spectrum with our best fit model and using the \cite{Fitzpatrick1999}
reddening law we find \mbox{$A_V=3.57\pm0.16$\,mag}, which is slightly higher than
the value (\mbox{$A_V=2.96\pm0.34$}\,mag) reported by \cite{Rodriguezetal2001}.\\
Furthermore, we discovered zeropoint shifts in the wavelengths calibration
of the OSIRIS spectra up to 54\,km/s (Table~\ref{tab:obs}). Correcting for
these and using Voigt profiles instead of Gaussian profiles in the RV fitting, our
results for \Ionw{He}{2}{5412} agree with the values reported by SG+15, but
for all other \ion{He}{II} lines we end up with conflicting RV amplitudes.\\
This issue was resolved by the realization that the spectra, and most notably
three of the four double-lined He\,II lines, are contaminated by
DIBs. Including the DIBs in the RV fitting, we obtain consistent results for
all four \ion{He}{II} lines and importantly, very distinct RV amplitudes of
$K_1=137\pm12$\,km/s for the primary and $K_2=214\pm14$\,km/s for the
secondary (using \Ionw{He}{2}{4542}, the only line not blended with any DIB).
These values no longer agree with the results of SG+15.\\
We then performed spectroscopic fits to the \ion{He}{II} lines using metal-free
non-LTE models. Using the results from the RV and spectral analysis, we
carried out light curve fits to the B-band, Sloan i-band, and ATLAS c-band
filters, to derive the geometry of the system. We find the effective
temperatures of both stars are about the same (\Teff$\approx40$\,kK), but higher than reported by
SG+15. The radii of the two stars (0.603\,\Rsol\ for the primary and 0.514\,\Rsol\ for the
secondary) are also found to differ from the results of SG+15, who found
$R=0.68$\,\Rsol\ for both stars. The inclination angle found by us
($i=63.59\pm0.54$\textdegree) agrees within the error limits with what is
reported by SG+15 ($i=64.7\pm1.4$\textdegree).\\
The most striking result of our analysis is that the mass ratio of the system
no longer equals one and that the dynamical masses of both stars
($M_1=0.66\pm0.11$\,\Msol\ and $M_2=0.42\pm0.07$\,\Msol) are significantly
smaller compared to the results of SG+15 ($M_1=M_2=0.88\pm0.13$\,\Msol).
The total mass of the system ($M=1.08\pm0.18$) no longer exceeds the 
Chandrasekhar mass limit, which again, is mainly a result of blends of
\Ionw{He}{2}{5412} with DIBs, which have led to an
overestimation of the dynamical masses of \hen by SG+15.
With these new findings, the merging event of \hen will not be recognised as
SN\,Ia, but most likely lead to the formation of a H-deficient star.\\
Based on the dynamical masses and atmospheric parameters revealed by our work,
we propose that the primary is a He-shell burning post-early AGB star,
and the secondary is the reheated core of a post-RGB star. The formation of
the system could be explained by a first stable mass transfer epsiode
in which \mbox{\mbox{Star\,2}} (now secondary) tranfered most of its mass to 
\mbox{\mbox{Star\,1}} (now primary) before it ignited He-core burning.
As \mbox{\mbox{Star\,1}} evolved up the early AGB, a common envelope was formed, and later
ejected, with the ejected material being now visible as the PN.\\
Even though the system can no longer be considered as a SN\,Ia progenitor, this
does not diminish the importance of \hen for studying common envelope evolution,
the formation of H-deficient stars via the double white dwarf merger channel,
and the creation of (asymmetrical) PNe via non-canonical (i.e., non-post-AGB)
evolutionary path ways. 
\hen is the only double-degenerate CSPN observed in an over-contact
configuration, thus, it might provide insights on the common envelope
ejection efficiency. Future spectroscopic observations offering a better S/N
especially in the blue part of the spectrum could improve the dynamical masses
and help to better constrain the evolutionary status of this interesting system.
A nebular abundance analysis will help to determine the metallicity
of the system. Finally, detailed evolutionary calculations that are able to reproduce
the history and future evolution of the system are highly encouraged.

\begin{acknowledgements}
      We thank M\'{o}nica Rodr\'{i}guez for providing us with the INT/IDS
      spectra. We appriciate useful discussions with David Jones and Tom Marsh
      during the CWDB meeting.       
      V.S. is supported by the \emph{Deut\-sche
        For\-schungs\-ge\-mein\-schaft, DFG\/} through grant GE 2506/9-1.
      Part of this work was supported by a MinCyT-DAAD bilateral cooperation
      program through grant DA/16/07.
      Based on data from the GTC PublicArchive at CAB (INTA-CSIC).
      Based  on  observations  collected  at  the  European  Organisation for
      Astronomical Research in the Southern Hemisphere under ESO programme 295.D-5032(A).
      IRAF is distributed by the National Optical Astronomy Observatory, which
      is operated by the Association of Universities for Research in Astronomy
      (AURA) under a cooperative agreement with the National Science
      Foundation. This work includes data from the Asteroid Terrestrial-impact
      Last Alert System (ATLAS) project. ATLAS is primarily funded to search
      for near earth asteroids through NASA grants NN12AR55G, 80NSSC18K0284,
      and 80NSSC18K1575; byproducts of the NEO search include images and
      catalogs from the survey area. The ATLAS science products have been made
      possible through the contributions of the University of Hawaii Institute
      for Astronomy, the Queen's University Belfast, the Space Telescope
      Science Institute, and the South African Astronomical Observatory. 
      
\end{acknowledgements}

\bibliographystyle{aa}
\bibliography{Hen2-428} 

\newpage

\begin{appendix}

\section{Figure}  
\begin{figure*}
\includegraphics[width=\textwidth]{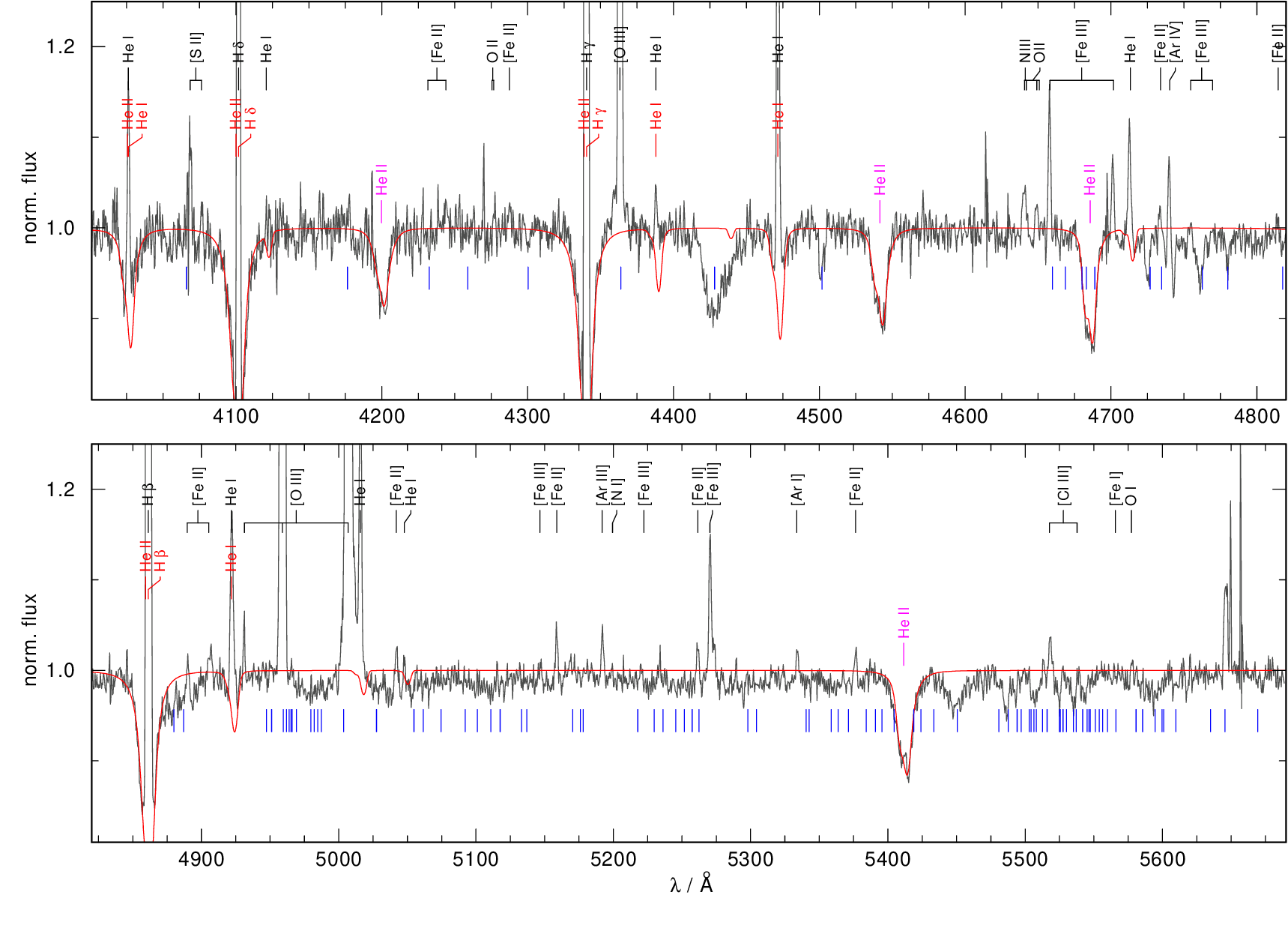}
  \caption{Normalized OSIRIS observation \#2 (gray) compared to our best fit TMAP model
    (red). The locations of known  diffuse interstellar bands (blue), nebular (black), and photospheric lines
  (red) are marked. Photospheric lines used in for the RV analysis are marked in magenta.}
  \label{fig:Osiris2bestfit}
\end{figure*}

\newpage
  
\section{Table}  
   
\begin{table}
\caption{Best lightcurve solution of \hen}
\label{lc}
\begin{tabular}{lcl}
\hline
\noalign{\smallskip}
\multicolumn{3}{l}{Fixed parameters:}\\
\noalign{\smallskip}
\hline
\noalign{\smallskip}
$q\,(=M_{2}/M_{1})$ & & $0.64$\\
$A_1^a$&&\multicolumn{1}{l}{1.0}\\
$A_2^a$ & & \multicolumn{1}{l}{1.0}\\
$g_1^b$&&\multicolumn{1}{l}{1.0}\\
$g_2^b$&&\multicolumn{1}{l}{1.0}\\
$x_1(B)^c$&&\multicolumn{1}{l}{0.25}\\
$x_1(i)^c$&&\multicolumn{1}{l}{0.17}\\
$x_1(c)^c$&&\multicolumn{1}{l}{0.20}\\
$x_2(B)^c$&&\multicolumn{1}{l}{0.25}\\
$x_2(i)^c$&&\multicolumn{1}{l}{0.17}\\
$x_2(c)^c$&&\multicolumn{1}{l}{0.20}\\
\noalign{\smallskip}
\hline
\noalign{\smallskip}
\multicolumn{3}{l}{Adjusted parameters:}\\
\noalign{\smallskip}
\hline
\noalign{\smallskip}
$i$ & [$^{\rm \circ}$] & $63.59\pm0.54$ \\
$T_{\rm eff}(1)$ & [K]& $40179\pm370$\\
$T_{\rm eff}(2)$ & [K]& $40356 \pm 175$\\
$\delta_1^d$&&$0.02174 \pm 0.0052$\\
$\delta_2^d$&&$0.0033 \pm 0.0021$\\
$\frac{L_1}{L_1+L_2}(B)^e$&&$0.5837 \pm  0.0068$\\
$\frac{L_1}{L_1+L_2}(i)^e$&&$0.5842 \pm  0.0068$\\
$\frac{L_1}{L_1+L_2}(c)^e$&&$0.5939 \pm  0.0168$\\
$\Omega_1^f$&&$2.965 \pm 0.017$\\
$\Omega_2^f$&&$2.965 \pm 0.017$\\
$l_3^g(B)$&&$0.0079 \pm 0.0043$\\
$l_3^g(i)$&&$0.0279 \pm 0.0066$\\
$l_3^g(c)$&&$0.2046 \pm 0.0312$\\
\noalign{\smallskip}
\hline
\noalign{\smallskip}
\multicolumn{3}{l}{Roche radii$^h$:}\\
\noalign{\smallskip}
\hline
\noalign{\smallskip}
$r_1$(mean)&[a]&$0.4472 \pm 0.0041$\\
$r_1$(pole)&[a]&$0.4121 \pm 0.0030$\\
$r_1$(point)&[a]&$-1.0000 $\\
$r_1$(side)&[a]&$0.4407 \pm 0.0039  $\\
$r_1$(back)&[a]&$0.4806 \pm 0.0056 $\\
\noalign{\smallskip}
$r_2$(mean)&[a]&$0.3809 \pm 0.0042$\\
$r_2$(pole)&[a]&$0.3465 \pm 0.0028 $\\
$r_2$(point)&[a]&$-1.0000 $\\
$r_2$(side)&[a]&$0.3676 \pm 0.0036 $\\
$r_2$(back)&[a]&$0.4212 \pm 0.0067
 $\\
\noalign{\smallskip}
\hline
\end{tabular}\\
\tablefoot{\\
$^{a}$ Bolometric albedo\\
$^{b}$ Gravitational darkening exponent\\
$^{c}$ Linear limb darkening coefficient; taken from \citet{ClaretBloemen2011} \\
$^{d}$ Radiation pressure parameter, see \citet{Drechsel1995}\\
$^{e}$ Relative luminosity; $L_2$ is not independently adjusted, but recomputed from $r_2$ and $T_{\rm eff}$(2)\\
$^{f}$ Roche potentials\\
$^{g}$ Fraction of third light at maximum\\
$^{h}$ Fractional Roche radii in units of separation of mass centers}
\label{tab:light_full}
\end{table}

\end{appendix}

\end{document}